# Constructing a Galaxy Cluster Catalog in IllustrisTNG-300 using the Mulguisin Algorithm

Lael Shin,[1] Jubee Sohn,[1,2] Young Ju,[3,4] Inkyu Park,[3,4] and Cristiano G. Sabiu[3,4]

[1]*Department of Physics & Astronomy, Seoul National University, Seoul 08826, Republic of Korea*
[2]*SNU Astronomy Research Center, Seoul National University, Seoul 08826, Republic of Korea*
[3]*Department of Physics, University of Seoul, Seoul 02504, Republic of Korea*
[4]*Natural Science Research Institute, University of Seoul, Seoul 02504, Republic of Korea*

## ABSTRACT

We present a new simulated galaxy cluster catalog based on the IllustrisTNG simulation. We use the Mulguisin (MGS) algorithm to identify galaxy overdensities. Our cluster identification differs from the previous FoF cluster identification in two aspects; 1) we identify cluster halos based on the galaxy subhalos instead of unobservable dark matter particles, and 2) we use the MGS algorithm that separates galaxy overdensities hosted by massive galaxies. Our approach provides a cluster catalog constructed similar to the observed cluster catalogs using spectroscopic surveys. The MGS cluster catalog lists 303 halos with $M_{200} > 10^{14}\,M_\odot$, including $\sim 10\%$ more than the FoF. The MGS catalog includes more systems because we separate some independent massive MGS cluster halos that are bundled into a single FoF algorithm. These independent MGS halos are apparently distinguishable in galaxy spatial distribution and the phase-space diagram. Because we constructed a refined cluster catalog that identifies local galaxy overdensities, we evaluate the effect of MGS clusters on the evolution of galaxies better than using the FoF cluster catalog. The MGS halo identification also enables effective identifications of merging clusters by selecting systems with neighboring galaxy overdensities. We thus highlight that the MGS cluster catalog is a useful tool for studying clusters in cosmological simulations and for comparing with the observed cluster samples.

## 1. INTRODUCTION

Constructing galaxy cluster catalogs is an essential foundation for systematic studies of the galaxy clusters and galaxies therein. Because galaxy clusters are the most massive gravitationally bounded system in the universe, the mass function of galaxy clusters are crucial probes to test cosmological models (e.g., White et al. 1993; Bahcall & Fan 1998; Wen et al. 2010; Allen et al. 2011; Papageorgiou et al. 2024). The dense environment in galaxy clusters provides a critical clue to studying how the environment regulates the growth of the cluster galaxies (e.g., Dressler 1980; Park et al. 2007; Peng et al. 2010; Ebeling et al. 2014; Gupta et al. 2017; Donnari et al. 2021).

Decades-long observations have constructed galaxy cluster catalogs based on various techniques. For example, detecting X-ray emission (e.g., Edge et al. 1990; Ebeling et al. 1998, 2010; Klein et al. 2023) and the distortion of the cosmic microwave background radiation (e.g., Melin et al. 2006; Bleem et al. 2015; Melin et al. 2021) due to the hot intracluster medium reveals large galaxy cluster samples. The most intuitive and long-standing technique is to identify galaxy overdensities based on photometric and spectroscopic surveys (e.g., Huchra & Geller 1982; Colless et al. 2001; Berlind et al. 2006; Robotham et al. 2011; Tempel et al. 2014; Rozo et al. 2015; Rykoff et al. 2016; Sohn et al. 2021). In particular, the cluster identification based on spectroscopy enables robust membership determination, and thus, the resulting cluster catalogs are well refined compared to other cluster catalogs (e.g., Sohn et al. 2018a,b; Clerc et al. 2020; Kirkpatrick et al. 2021).

Cosmological simulations provide opportunities to compare the study of observed clusters and cluster galaxies with the theoretical predictions based on our understanding of current galaxy and structure formation models. There are many galaxy cluster catalogs constructed based on numerical simulations (e.g., Ghigna et al. 1998; Dolag et al. 2009; Schaye et al. 2015; ZuHone et al. 2018). Many of these cluster catalogs are built based on the simulated particles (e.g., Schaye et al. 2015; Barnes et al. 2017; Springel et al. 2018; Pillepich et al.

2018a). For example, the galaxy cluster catalog for IllustrisTNG (Springel et al. 2018; Pillepich et al. 2018a), one of the largest cosmological hydrodynamic simulations, is constructed based on the standard Friends-of-Friends (FoF) algorithm. The FoF algorithm is applied to dark matter particles to find the clusters, and the baryonic particles are later assigned to the same FoF halo as their closest dark matter particle (Springel et al. 2018; Pillepich et al. 2018a,b). Because these simulated particles (particularly the dark matter particles) are not directly observable, the resulting simulated cluster catalogs are not directly comparable with the observed cluster catalogs, that are commonly based on galaxies. Thus, constructing a new simulated galaxy cluster catalog based on the same manner we applied to observations is necessary for a fair comparison to test and improve the understanding of galaxy clusters.

Here, we build a new cluster catalog based on IllustrisTNG using a similar approach to observed cluster identification. We select the galaxy subhalos in the cosmological simulations, and we identify clusters of galaxies. For grouping, we use an up-to-date `Mulguisin` algorithm (Ju et al. 2023). The Mulguisin (MGS) algorithm basically bundles galaxies (subhalos) into clusters when galaxies are within the linking length, similar to the conventional FoF algorithm. In addition to the positional information, the MGS algorithm uses the mass of galaxies to distinguish overdensities that are hosted by different massive galaxies. Ju et al. (2023) highlight that the MGS algorithm identifies simulated clusters very similar to human visual identification.

In Section 2, we first introduce the IllustrisTNG simulation used to build the MGS cluster catalog. We describe the MGS algorithm in details in Section 3. We also explain the MGS cluster catalog from IllustrisTNG. We then explore the properties and strength of the MGS cluster catalog by comparing it with the previous simulated cluster catalogs in Section 4. We demonstrate two examples where the new MGS cluster catalog is beneficial to studying cluster galaxy properties and identifying merging cluster candidates in Section 5. We conclude in Section 6. We use the Planck 2015 cosmological parameters (Planck Collaboration et al. 2016), with $\Omega_{\rm m} = 0.3089$, $\Omega_{\Lambda} = 0.6911$, and $H_0 = 67.74 \rm\,km\,s^{-1}\,Mpc^{-1}$.

## 2. DATA

We construct a new cluster catalog based on IllustrisTNG, a suite of magnetohydrodynamical cosmological simulations performed with a quasi-Lagrangian code `AREPO` (Marinacci et al. 2018; Naiman et al. 2018; Nelson et al. 2018; Pillepich et al. 2018a; Springel et al. 2018). IllustrisTNG succeeds the previous Illustris project (Vogelsberger et al. 2014a,b) with larger volumes and better resolutions. IllustrisTNG also implements an improved galaxy evolution model, such as black hole–driven feedback and stellar population evolution (Weinberger et al. 2017; Pillepich et al. 2018b).

We use IllustrisTNG300, a set of simulations with the box size of $\sim 300$ Mpc side, to identify a large number of massive cluster halos. TNG300-1 is the highest-resolution simulation among IllutrisTNG300 with a baryon mass resolution of $\sim 1.1 \times 10^7 \, {\rm M_\odot}$, and a resolution for the dark matter of $\sim 5.9 \times 10^7 \, {\rm M_\odot}$. The high resolution enables robust analysis of structures within a wide mass range from low-mass galaxies with the stellar mass of $\sim 10^9 \, {\rm M_\odot}$ to galaxy cluster with mass scales of $\sim 10^{14} \, {\rm M_\odot}$. IllustrisTNG provides a halo catalog for each simulation built based on the Friends-of-Friends (FoF) algorithm (Huchra & Geller 1982; Davis et al. 1985). The FoF algorithm identifies the TNG FoF halos using dark matter particles. The FoF algorithm requires a linking length to bundle the particles within a single system. The linking length used for the TNG FoF halos is $b = 0.2 \, [d_{\rm mean}]$, where $d_{\rm mean}$ is the mean separation of dark matter particles. The algorithm assigns the other types of particles (including gas, star, and black holes) to the halos where their nearest dark matter particles belong.

The TNG FoF halos are often compared with observed galaxy structures. However, the TNG FoF halo identification is based on dark matter particles that are not directly observable. For a fair comparison with observed groups and clusters, the TNG systems that are similarly identified with the observed techniques would be useful.

We, therefore, build a new simulated cluster catalog based on subhalos, corresponding to the galaxies in the observations, instead of using the dark matter particles. We use TNG subhalo catalog constructed based on the `SUBFIND` algorithm (Springel et al. 2001; Dolag et al. 2009). We select 276,007 subhalos in TNG300-1 with cosmological origin (i.e., *SubhaloFlag* = 1) and stellar mass larger than $10^9 \, {\rm M_\odot}$. This stellar mass selection corresponds to the stellar mass limit of dense spectroscopic surveys of massive clusters (Sohn et al. 2017, 2022).

## 3. THE MULGUISIN HALO CATALOG

Our primary goal is to construct a new simulated galaxy cluster catalog directly comparable to the observed cluster catalogs. We introduce the MGS algorithm we used for building the halo catalog in Section 3.1. We describe the construction of the MGS cluster catalog in Section 3.2. We explore the physical properties of the MGS halos in Section 3.3.

### 3.1. The Mulguisin Algorithm

The Mulguisin (MGS) algorithm (Ju et al. 2023) was initially devised as a jet detector for the Large Hadron Collider (LHC) in the ATLAS Collaboration (Bosman et al. 1998). Mulguisin is a Korean word meaning a ghost living in lakes. The main idea of the MGS algorithm is that we can figure out where Mulguisins hide in lakes by gradually draining water from the lake. A Mulguisin near the surface will appear soon after the water drainage, while another Mulguisin deep in the water will appear when most water is drained. Based on this idea, the MGS algorithm can identify the spatial distribution of particle overdensities using an analogy between water depth and particle density.

We use the MGS algorithm to identify overdensities of subhalos in TNG300. The MGS algorithm first sorts subhalos in the mass (or local density) order. We start by identifying the most massive subhalo in TNG300 as a cluster seed. We then move on to the next most massive subhalo. We compute its distance from all other subhalos more massive than this subhalo; for this second massive subhalo, we only compute the distance to the most massive subhalo in TNG300 (the cluster seed). If this target subhalo has more massive neighboring subhalos within a linking length, the MGS algorithm classifies those two subhalos as a single system. If not, the target subhalo becomes a new cluster seed. We repeat this procedure until we assign a cluster membership for every subhalo. In some cases, the target subhalos may have multiple more massive subhalos within a linking length. We group the target subhalo as a member of the system where the nearest massive subhalo belongs.

### 3.2. The Mulguisin Halo Identification

We apply the MGS algorithm to subhalos in TNG300. Identifying cluster halos based on subhalos enables a direct comparison with observed cluster catalogs. To avoid mis-identification of halos at the edges of simulation boxes, we consider the periodic boundary condition, similar to the FoF halo identification.

The MGS algorithm sorts the subhalos to identify cluster seeds based on stellar mass or galaxy number density. Here, we sort galaxies with stellar mass. In the case of the MGS catalog based on the density sorting, the follow-up comparison with observation is not straightforward because measuring the local number density of the galaxy is often challenging.

The MGS algorithm requires a single parameter, the linking length. We empirically determine the linking length for the MGS algorithm that similarly reproduces the total mass density distribution of the TNG FoF halos. Because our primary goal is to construct a cluster-like halo catalog, we use the TNG FoF halos with total masses larger than $10^{14}$ M$_\odot$ (hereafter, FoF cluster halos). For the empirical test, we first compute the mean separation of 276,007 TNG subhalos; $d_{\rm mean} \sim 4.6$ Mpc. We then identify MGS halos with various linking lengths ranging $0.1 \leq b\,[d_{\rm mean}] \leq 0.5$. Here, b indicates the normalized linking length in the unit of $d_{\rm mean}$.

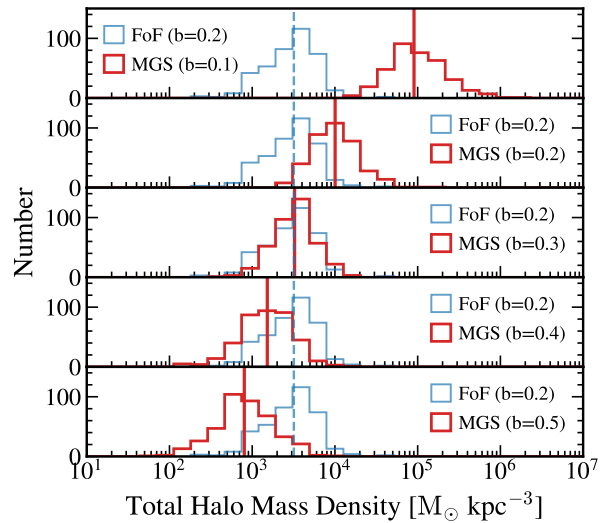

**Figure 1.** Mass density distributions of MGS (red) and FoF (blue) halos with the sum of the subhalo masses $> 10^{14}$ M$_\odot$. The sum is carried out within a spherical volume with a boundary to the farthest member subhalo from the halo center. The linking length b for the MGS algorithm in units of the mean subhalo separation varies from 0.1 to 0.5 from top to bottom panels. Vertical lines show the median of the distributions.

Figure 1 shows the total mass density distributions of FoF cluster halos (blue histograms) with the fixed linking length $b = 0.2$. To compute the total mass density, we first set the spherical halo boundary with the radius of the distance to the farthest member subhalo from the halo center. We then select member subhalos identified by each algorithm within the spherical volume. The total halo mass indicates the sum of the total mass of member subhalos (i.e., *SubhaloMass* from the IllustrisTNG group catalog), and the total mass of the subhalo indicates the sum of dark matter, stars, gas, and black hole particles bound to the subhalo.

The red histograms of Figure 1 display the total halo mass density of MGS halos identified with various linking lengths (from top to bottom panels). Similar to the FoF cluster halos, we only select the MGS halos with total masses larger than $10^{14}$ M$_\odot$. The mass density of MGS halos generally decreases as the linking length increases. With the larger linking lengths, the MGS al-

4gorithm includes subhalos distant from the halo center. Thus, the volume covered by halos increases significantly as the linking length increases. However, the number of subhalos within the increased volume is small. In other words, the mass increment is less significant than the volume increment. Therefore, the total mass density distribution decreases for the MGS systems identified by larger linking lengths.

Figure 1 demonstrates that the MGS halos identified with the linking length $b = 0.3$ show essentially identical total mass density distribution of the FoF cluster halos. The median total mass density of these MGS halos is $\sim 3.3 \times 10^3 \, M_\odot \, \text{kpc}^{-3}$, roughly corresponding to $\sim 26$ times the critical density of the universe, indicating that they are cluster-like halos (More et al. 2011; Duarte & Mamon 2014). Hereafter, we use the MGS algorithm with $b = 0.3$ to identify halos within TNG300.

### 3.3. *The IllustrisTNG MGS Cluster Halo Catalog*

We identify 124,155 halos based on the MGS algorithm with the linking length $b = 0.3$. We apply the algorithm to 276,007 subhalos with the stellar mass $> 10^9 \, M_\odot$ in the $z = 0$ snapshot of TNG300. The resulting number of MGS halos is much smaller than the 17,625,892 TNG FoF halos due to the different basis for each halo identification (i.e., subhalos versus. dark matter particles).

We estimate $M_{200}$ and $R_{200}$ for 3,082 MGS halos with 10 or more cluster members. $M_{200}$ is the sum of the mass of dark matter, star, gas, and blackhole particles within a sphere where the enclosed mass density is 200 times the critical density at the redshift (i.e., $z = 0$); $R_{200}$ is the radius of the sphere. Figure 2(a) shows the number density distribution of the MGS halos as a function of $M_{200}$. There are 303 MGS halos with $M_{200}$ larger than $10^{14} \, M_\odot$; hereafter we refer to them as MGS cluster halos.

We also compute the cluster velocity dispersion ($\sigma_{\text{cl}}$) to explore the $M_{200}$ - $\sigma_{\text{cl}}$ relation of MGS halos. We first measure the projected cluster-centric distance of member subhalos, $R_{\text{cl}} = \sqrt{\Delta X^2 + \Delta Y^2}$, where $\Delta X$ and $\Delta Y$ are the distance of the cluster members from the cluster center. For the member subhalos within $R_{\text{cl}} < R_{200}$, we calculate the 1D velocity dispersion along with $z-$axis using the biweight technique (Beers et al. 1990). We estimate the velocity dispersion uncertainty from the $1\sigma$ standard deviation of 1000 bootstrap resamplings. The cluster velocity dispersion we compute corresponds to the line-of-sight (LOS) velocity dispersion from observations (e.g., see Section 2.2. in Sohn et al. 2022).

In Figure 2(b), we show the relation between $M_{200}$ and the cluster velocity dispersion of MGS halos. We derive the best-fit relation, $\sigma_{\text{cl}} \propto M_{200}^\alpha$, using `scipy.curve_fit` (Virtanen et al. 2020). The best-fit relation from MGS halos has a slope $\alpha = 0.358 \pm 0.012$, consistent with the slopes from previous works (e.g., $\alpha = 0.339$; Marini et al. 2021, $\alpha = 0.343$; Sohn et al. 2022).

Table 1 lists the properties of 303 MGS cluster halos in TNG300 including BCG SubfindID, $M_{200}$, $R_{200}$, and $\sigma_{\text{cl}}$, $N_{\text{member}}$. Table 2 includes the properties of member subhalos of the MGS cluster halos: host MGS halo ID, SubfindID, position, velocity, stellar mass, and the mean stellar age.

## 4. COMPARISON WITH PREVIOUS SIMULATED CLUSTER CATALOGS

We explore the properties of the MGS halo catalog by comparing it with simulated galaxy cluster catalogs from previous studies. We first examine the halo mass function and the M - $\sigma_{\text{cl}}$ relation of MGS and original TNG FoF halos in Section 4.1.1. We then investigate the distinct MGS structure identification in Section 4.1.2. We also identify merging cluster candidates in Section 4.2.

### 4.1. *Comparison with IllustrisTNG FoF cluster catalog*
#### 4.1.1. *Halo mass function & $M_{200}$ - $\sigma_{cl}$ relation*

Figure 2(a) compares the mass function of the MGS (red circles) and FoF (blue crosses) halos. The MGS halo mass function is generally consistent with the mass function of FoF halos. However, there are slightly more MGS halos in the most massive bins ($M_{200} > 10^{14} \, M_\odot$). At $M_{200} > 10^{14} \, M_\odot$, the MGS algorithm identifies 303 halos, while the FoF algorithm identifies 280 halos. The number of halos differs because the MGS algorithm separates some massive structures identified as a single FoF halo (see more details in Section 4.1.2).

Figure 2(b) shows the relation between $M_{200}$ and the $\sigma_{cl}$ of MGS halos. The solid and dashed lines indicate the best-fit M - $\sigma_{cl}$ relation for MGS and FoF halos. We derive the best-fit relations:

$$\log \sigma_{\text{cl,MGS}} = (0.358 \pm 0.012) \log M_{200} + (-2.398 \pm 0.165) \quad (1)$$

and

$$\log \sigma_{\text{cl,FoF}} = (0.333 \pm 0.011) \log M_{200} + (-2.046 \pm 0.166) \quad (2)$$

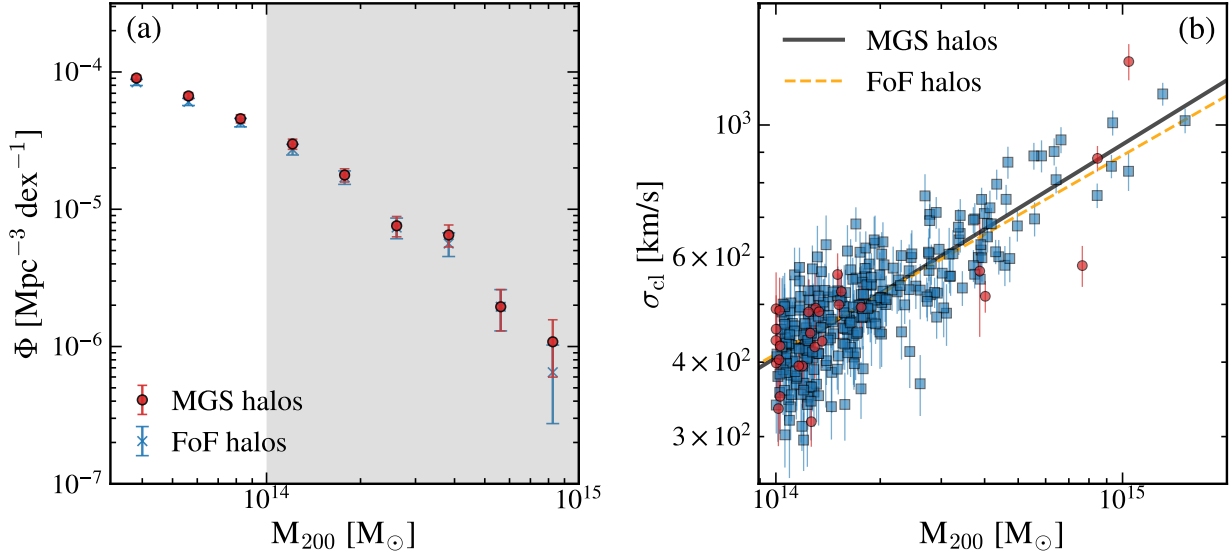

**Figure 2.** (a) Halo mass functions of FoF (blue) and MGS halos (red). The MGS mass function slightly exceeds the FoF mass function at the most massive bins. The systems in the gray shaded area are the cluster halos with $M_{200} > 10^{14}\,M_\odot$ (b) $M_{200}$-$\sigma_{cl}$ relation of the MGS halos with $M_{200} > 10^{14}\,M_\odot$. The blue squares indicate the MGS halos coincident with the original FoF halo catalog, and the red dots are the newly identified massive MGS halos. The solid line shows the best-fit M-$\sigma$ relation of MGS halos, and the dashed line indicates the relation of the original FoF halos.

**Table 1.** MGS Clusters in IllustrisTNG-300

| MGS ID | BCG Subfind ID[a] | Host FoF ID[b] | log $M_{200}$ ($M_\odot$) | $R_{200}$ (kpc) | $\sigma_{cl}$[c] (km s$^{-1}$) | $N_{member}$[d] |
|---|---|---|---|---|---|---|
| 0 | 0 | 0 | 15.2 | 2422 | 1031 ± 49 | 334 |
| 1 | 11748 | 1 | 15.1 | 2304 | 1142 ± 53 | 275 |
| 2 | 17908 | 2 | 15.0 | 2136 | 836 ± 61 | 128 |
| 3 | 27878 | 4 | 14.9 | 1993 | 803 ± 37 | 282 |
| 4 | 22736 | 3 | 15.0 | 2063 | 1016 ± 48 | 244 |

NOTE— [a] Subfind ID of the MGS cluster BCG.
[b] Host FoF halo ID of the BCG.
[c] Cluster velocity dispersion.
[d] Number of MGS cluster member subhalos.
(The entire table is available in machine-readable form.)

Considering the intrinsic scatter $\sim 0.06$ dex, the M - $\sigma_{cl}$ relation of MGS halos is overall consistent with the relation of FoF halos. Including the $\sim 20$ newly identified MGS halos with $M_{200}$ more massive than $10^{14}\,M_\odot$ does not significantly change the original M - $\sigma_{cl}$ relation.

#### 4.1.2. Structure identification

We next investigate the cases where a single FoF system consists of multiple massive MGS systems. Among 280 FoF systems with $M_{200} > 10^{14}\,M_\odot$, 40 systems include structures with more than 20 member subhalos ($N_{member}$) and with $M_{200} > 10^{13}\,M_\odot$ identified by the MGS algorithm. For 17 FoF systems, the MGS algorithm identifies massive multiple structures with $N_{member} > 20$ and with $M_{200} > 10^{14}\,M_\odot$. We select the systems with $N_{member} > 20$ to select the similar systems with the observed FoF clusters with a similar number of members (Sohn et al. 2021).

**Table 2.** MGS Cluster Member Subhalos

| Host MGS ID[a] | Subfind ID | x (kpc) | y (kpc) | z (kpc) | $v_x$ (km s$^{-1}$) | $v_y$ (km s$^{-1}$) | $v_z$ (km s$^{-1}$) | log $M_{star}$ ($M_\odot$) | Age$_{star}$[b] (Gyr) |
|---|---|---|---|---|---|---|---|---|---|
| 0 | 0 | 64959 | 72830 | 217521 | 472 | 451 | -261 | 12.9 | 11.52 |
| 0 | 2 | 65675 | 72472 | 218292 | 2022 | 1495 | -1797 | 11.9 | 11.42 |
| 0 | 4 | 65402 | 73265 | 218289 | -260 | -2222 | -564 | 11.5 | 11.18 |
| 0 | 32 | 65666 | 72874 | 218079 | -203 | 2810 | 562 | 11.0 | 9.68 |
| 0 | 38 | 65212 | 70224 | 218668 | 402 | 588 | 711 | 11.0 | 10.92 |

NOTE— [a] Identification number of the host MGS halo.
[b] Subhalo mean stellar age (see Section 5).
(The entire table is available in machine-readable form.)

We investigate further the 17 FoF systems with massive MGS structures. Figure 3 shows the spatial distribution of member subhalos of the 17 FoF halos with multiple massive MGS structures ($M_{200} > 10^{14}\,M_\odot$) except for the two FoF halos with too close substructures ($ID_{FoF} = 1$ and 2, see Section 4.2). To clarify the separation between structures, we choose the plane where the projected separation between the centers of the two massive MGS substructures is most distinctive. The black contour shows the number density of the FoF member subhalos. Circles mark the $R_{200}$ and the center of the massive MGS structures with $M_{200} > 10^{13}\,M_\odot$ and $N_{member} > 20$.

In Figure 3, the member density map of the FoF subhalos shows multiple substructures. The FoF algorithm bundles the massive substructures into a single system. However, the MGS algorithm clearly separates the overdensities by identifying the multiple peaks in the subhalo number density distribution. For the cases of $ID_{FoF} = 38$ and 46, the subhalos at the upper-left density peak are not marked with color because the identified MGS halos are less massive than the criteria $M_{200} > 10^{13}\,M_\odot$ and $N_{member} > 20$. The centers of the other massive MGS halos are consistent with the subhalo number density peaks.

Figure 4 displays the phase space diagrams of the 15 systems. Here, the phase-space diagram, often referred to as the R-v diagram, shows the line-of-sight peculiar velocity ($\Delta v$) of subhalos as a function of projected distance from the FoF centers. We choose the same plane as Figure 3 to measure the projected distance and the line-of-sight velocity. We plot members of different MGS halos with different colors.

In Figure 4, the MGS member distribution near the original FoF halo center shows a trumpet-like pattern observed in typical clusters (Kaiser 1987; Regos & Geller 1989). The other MGS structures, however, are clearly distinguished from the central MGS halo. The member subhalos in the other massive MGS halos extend to larger distances ($> R_{200}$), and the velocity dispersions are much larger than that expected from the trumpet-like pattern. The median velocity of member subhalos of some MGS halos (e.g., $ID_{FoF}=6, 7, 29, 83$) even shows a significant velocity offset ($> 1\sigma_{cl}$) from that for the central MGS halos.

### 4.2. Comparison with the merging cluster catalog

We next compare the MGS halo catalog with a merging cluster catalog from Łokas (2023). Łokas (2023) identified ten merging cluster candidates in IllustrisTNG-300 based on the visual inspection of the bow shock in the gas temperature map. The MGS algorithm identifies all of the ten systems, but the MGS algorithm separates substructures of only two merging clusters. The MGS algorithm identifies a single system for the other eight merging clusters in Łokas (2023).

Figure 5 shows the distances (the blue-filled histogram) between the most massive subhalos in merging systems identified by Łokas (2023). These merging clusters generally have a small separation $< 1$ Mpc. Because Łokas (2023) visually identified the bow shock within $R_{200}$, typically $\sim 1.1$ Mpc, they tend to identify merging systems with a very tight separation.

We then display a similar separation distribution with the MGS cluster catalog. We measure the distance between 303 MGS halo with $M_{200} > 10^{14}\,M_\odot$ and the nearest neighbor with $M_{200} > 10^{13}\,M_\odot$. The red histogram shows the minimum separation between the nearby MGS structures. With the linking length we used (i.e., $\sim 1$ Mpc), the MGS algorithm can separate the systems with a separation larger than 1 Mpc. Thus, the eight merging systems in Łokas (2023) that have smaller separations than the 1 Mpc limit cannot be identified as multiple structures with our approach. We further inves-tigate the possibility of identification of merging systems with the MGS algorithm in Section 5.2.

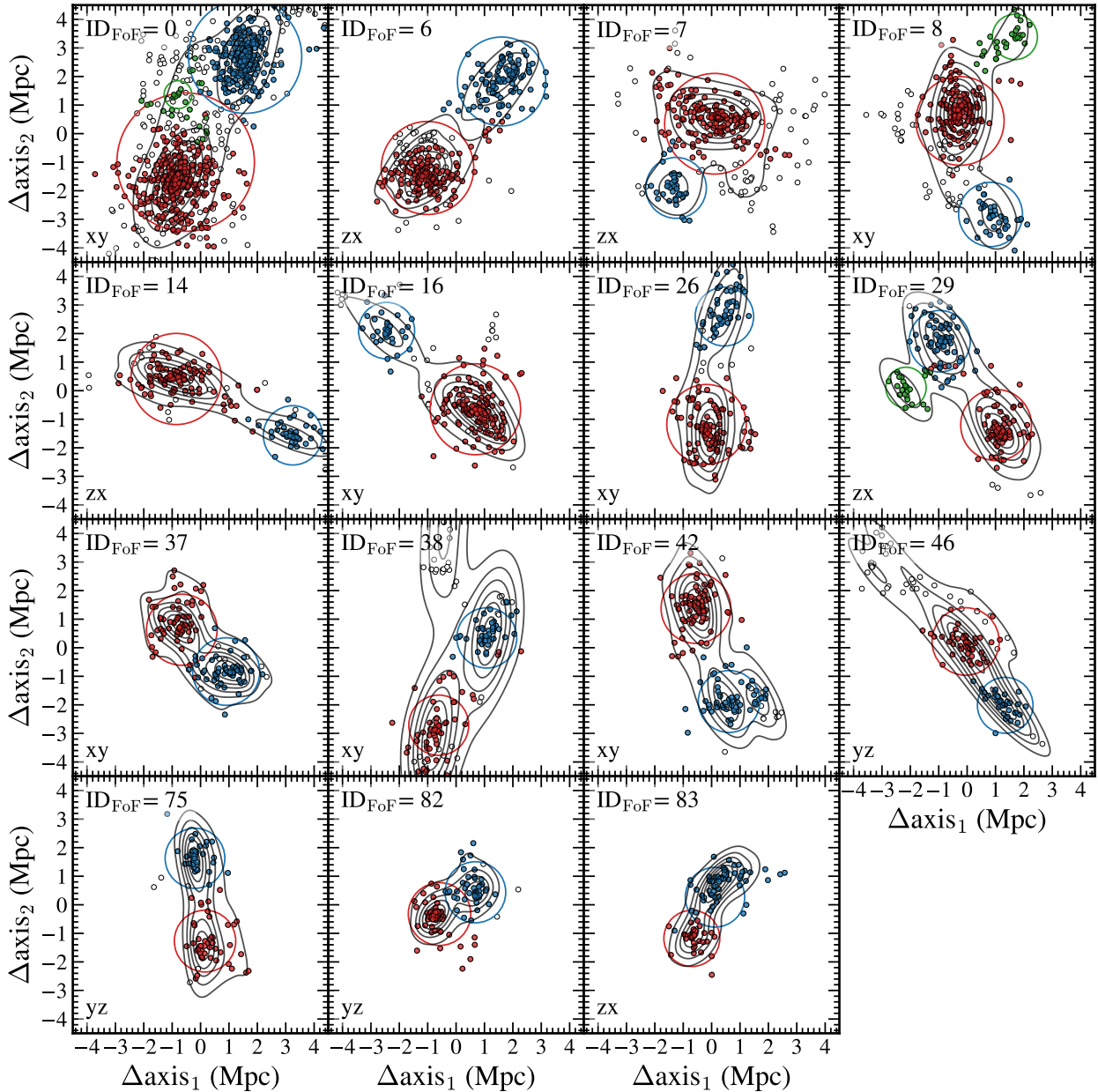

**Figure 3.** Example FoF halos with $M_{200} > 10^{14}\,M_\odot$, where MGS identifies multiple halos. Points show the spatial distribution of each FoF halo member subhalo. Black contours show their number density distributions. Circles mark the center and $R_{200}$ of halos we identify based on the MGS algorithm. Different colors indicate MGS halo memberships within the FoF halo. We show the distribution of subhalos on the plane (e.g., x-y, y-z, or x-z planes) where the separations of multiple structures are distinctive.

## 5. STRENGTH OF USING THE MULGUISIN HALO CATALOG

We construct an independent galaxy cluster catalog for IllustrisTNG based on the MGS algorithm and compare it with previous simulated galaxy cluster catalogs: the original IllustrisTNG FoF catalog and the merging cluster catalogs. Here, we demonstrate a few other examples where the new MGS catalog is beneficial to studying cluster galaxy properties. We examine the mean stellar age of subhalos and the fraction of quenched subhalos within the MGS halos. We also present merging MGS halo candidates with bow shocks in the gas temperature map.

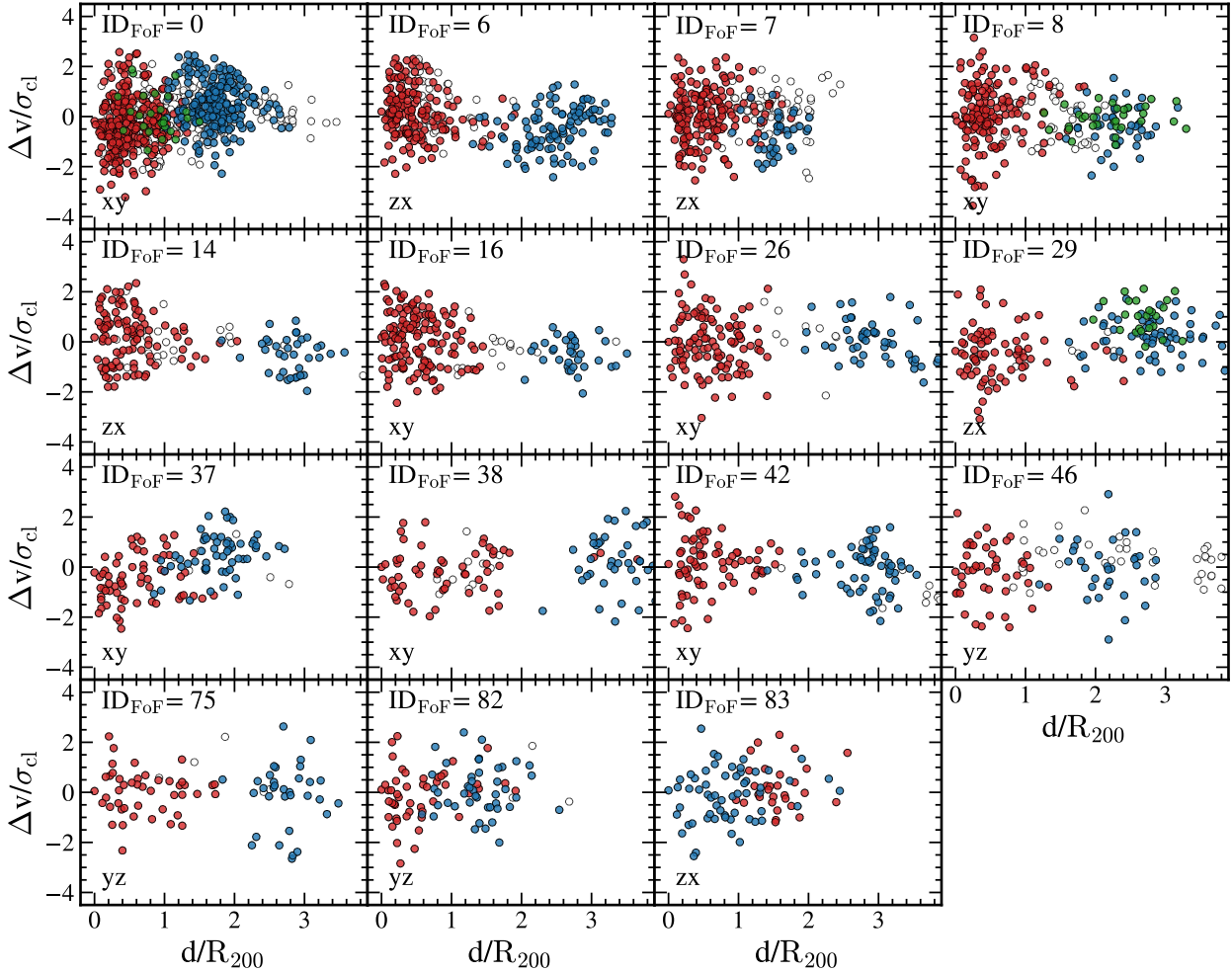

**Figure 4.** Phase space diagrams of massive halos in Figure 3. Points are the member subhalos with $M_* > 10^9$ $M_\odot$ within the FoF halos; different colors indicate memberships of the multiple MGS halos within the FoF halo. The members of different MGS halos are clearly separated on the phase-space diagrams.

### 5.1. *Subhalo mean stellar age & quenched fraction distribution*

We explore the subhalo mean stellar age distribution within the MGS clusters. We first measure the mean stellar age of the member subhalos in the MGS clusters. We collect the formation time (i.e., *GFM_StellarFormationTime* from the IllustrisTNG group catalog) of the member stellar particles of subhalos. We then compute the mass-weighted mean stellar age ($\overline{\text{Age}_*}$) of the subhalos.

The upper panels of Figure 6 display the distribution of the mean stellar age of subhalos. Figure 6(a) shows the mean stellar age of subhalos in the 17 FoF clusters with multiple MGS counterparts. We plot the median distribution of mean stellar age as a function of projected distance from the FoF center. Because we compile the data from 17 FoF systems, we normalize the projected distance from the FoF center (i.e., $d_{\text{proj}}$) with the distance between the BCGs of the two most massive MGS counterparts (i.e., $d_{\text{BCG1-BCG2}}$). We also normalize the mean stellar age with the mean stellar age of the subhalos near the FoF centers ($\overline{\text{Age}_{*,center}}$). We compute the median of the normalized mean stellar age at each bin. The error bar indicates the $1\sigma$ of the median values from the 1000 bootstrap resamplings.

The mean stellar age of the FoF member subhalos decreases until $d_{\text{proj}} \sim 0.6\,d_{\text{BCG1-BCG2}}$ as a function of projected distance, but it increases at $d_{\text{proj}}/d_{\text{BCG1-BCG2}} > 0.6$. Because the subhalos at $d_{\text{proj}}/d_{\text{BCG1-BCG2}} > 0.6$ are under the influence of the independent halos, their mean stellar ages increase as the subhalos are close to the BCGs of the independent massive halos.

Figure 6(b) displays the same relation but based on the MGS halos. We use 37 MGS halos within the same

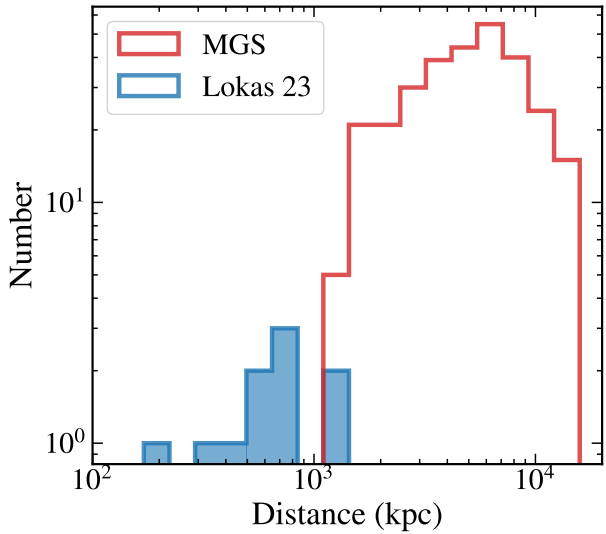

**Figure 5.** The distribution of the separation between the BCGs of the main cluster and subcluster of ten merging clusters from Łokas 2023 (blue). The distance between the BCGs of 303 MGS structures with $M_{200} > 10^{14} \, M_\odot$ and their nearest MGS neighbors with $M_{200} > 10^{13} \, M_\odot$ (red). Two clusters with BCG separation $\sim 1$ Mpc are overlapped. The separation of the other merging clusters from Łokas 2023 is smaller, and the distance between massive MGS structures is larger than 1 Mpc.

17 FoF halos we used for Figure 6(a). Here, we simply compute the normalized projected distance from the MGS halos. In contrast to Figure 6(a), the median normalized stellar age monotonically decreases as a function of projected distance. Because we identify independent MGS halos that are grouped as a single FoF halo, the MGS halo catalog provides a clearer sample to investigate the mean stellar age distribution without contamination by the surrounding massive structures.

Figure 6(c) and (d) show a similar comparison with the upper panels of Figure 6, but based on the quenched fraction. The quenched fraction indicates the number ratio between the quenched subhalos and all subhalos within the normalized projected distance bins. Here, the quenched subhalos have the star formation rate (SFR) 10 times lower than the main star-forming sequence at a given stellar mass. We define the main star-forming sequence (i.e., SFR - $M_*$) based on subhalos with $10^9 < (M_*/M_\odot) < 10^{11}$ in TNG300 following Bluck et al. (2016); Donnari et al. (2019); Wang et al. (2021). The main star-forming sequence is:

$$SFR = (0.78 \pm 0.07) \left( \frac{M_*}{M_\odot} \right) + (-7.9 \pm 0.7) \, M_\odot \, yr^{-1}. \tag{3}$$

The quenched fraction is generally high ($> 60\%$) for both FoF and MGS subhalos. Similar to the mean stellar age distributions, the quenched fraction of the FoF halos increases slightly near the center of secondary BCGs. However, the quenched fraction of the MGS subhalos decreases continuously as a function of the projected distance.

Decreasing the mean stellar age of subhalos and the quenched fraction as a function of clustercentric distance in simulated cluster halos is consistent with observations. Many observational studies examine the environmental effect on galaxy evolution in clusters depending on the distance from the cluster center (e.g., Balogh et al. 1999; Deshev et al. 2017; Sohn et al. 2019; Oman et al. 2021; Coenda et al. 2022; Kim et al. 2023). For example, Sohn et al. (2019) investigate the $D_n4000$ (the mean stellar age indicator of a galaxy) of galaxies in a massive cluster A2029 based on extremely dense spectroscopy. They show that the median $D_n4000$ decreases generally as a function of clustercentric distance. They particularly demonstrate that the $D_n4000$ of member galaxies within the massive substructures of A2029 clearly depends on the distance from the center of substructures instead of the distance from the center of the main cluster.

Comparison between the age distributions of subhalos within the FoF and MGS halos indicates the advantage of the MGS halo catalog. As Sohn et al. (2019) studied the $D_n4000$ distribution based on the dense spectroscopy, the MGS halo catalog offers a robust sample to study galaxy evolution in clusters separating the effects of neighboring massive structures. Investigating the galaxy properties with the MGS halo catalogs enables a better evaluation on the impact of cluster halos compared to a similar study with the FoF catalog that does not separate the local structures.

### 5.2. Identification of merging cluster candidates with bow shocks

Another important application of the MGS cluster catalog is the merging system identification (see Section 4.2). The MGS algorithm distinguishes galaxy number density peaks with small separations. The refined cluster catalog with the MGS algorithm enables the identification of currently merging cluster candidates.

We first examine the 17 FoF systems we described in Section 5.1 to investigate if these systems with nearby MGS halos are in a merging phase. Following Łokas (2023), we identify merging systems based on the gas temperature map. Łokas (2023) identifies bow shock features visually in the gas temperature map constructed based on TNG gas components. In the merging cluster catalog they built, Łokas (2023) includes the two MGS systems ($ID_{MGS}$ = 1 and 2). However, none of remaining 15 systems in our sample was classified as merging clusters by Łokas (2023).

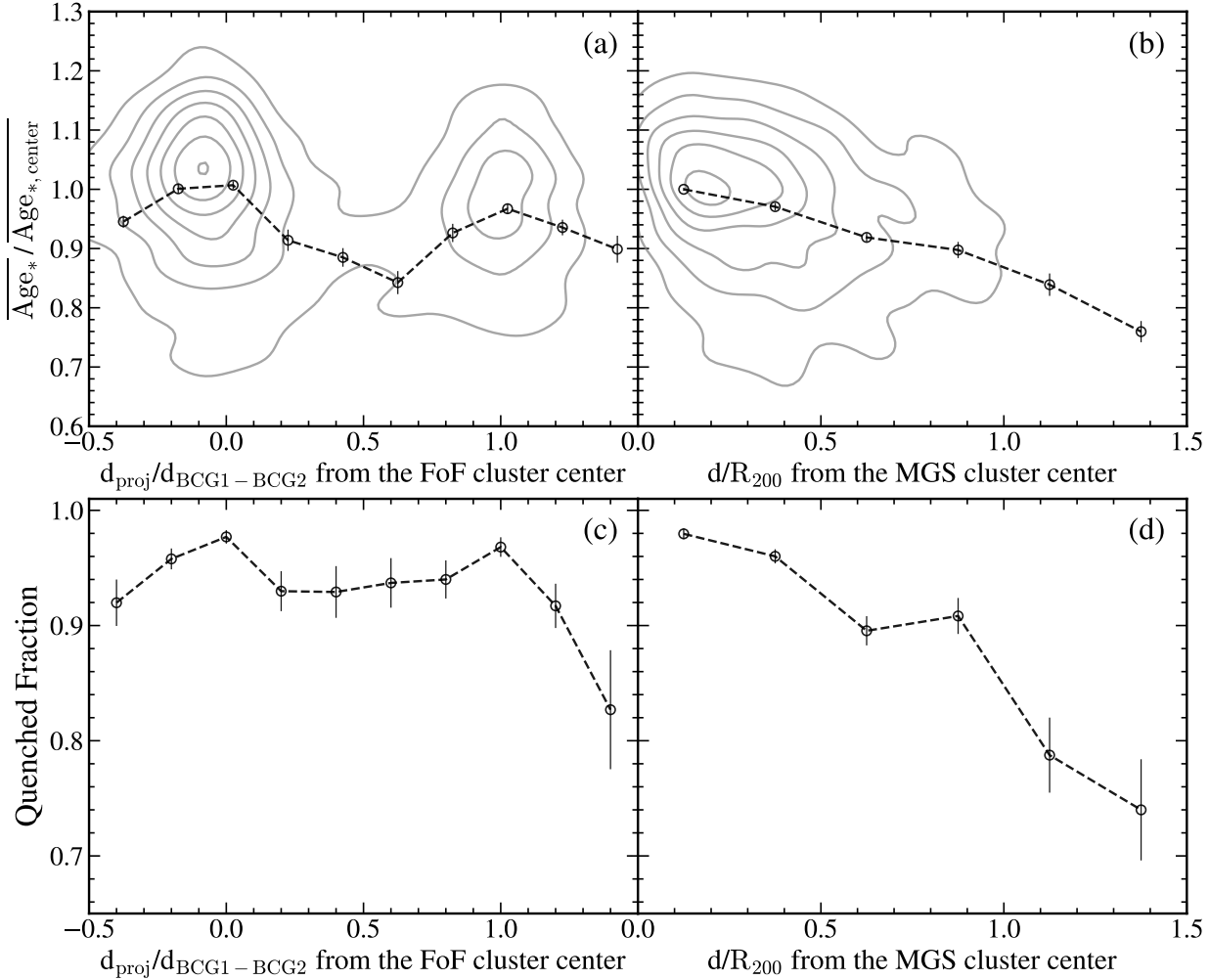

**Figure 6.** (a) The mean stellar age distribution (gray contour) of member subhalos of the 17 TNG FoF halos with multiple MGS counterparts (shown in Figure 3). The x-axis shows the 1D projected distance between member subhalos and the FoF BCG onto the axis connecting the two most massive MGS (sub)structure BCGs. We normalize the 1D projected distance with the distance between the two MGS BCGs. Circles mark the median distribution. Note that there is an upturn between two BCGs. (b) The same as panel (a), but based on the distance from the MGC BCGs normalized by the virial radius. Panels (c) and (d) display the quenched fraction distribution along with the clustercentric distance similar to panels (a) and (b), respectively.

From our visual inspection, we identify four systems ($ID_{MGS}$ = 1, 2, 97, 120) with the bow shock-like features in the gas temperature maps. The upper panels of Figure 8 display the gas temperature maps of the four systems. The contours with different colors show the density map of the independent MGS halos. For the two systems we identity ($ID_{MGS}$ = 97 and 120), we identify a weak but sharp temperature contrast between two MGS halos. These features become more obvious when we plot the shock-dissipated energy map (the lower panels of Figure 8). The shock-dissipated energy indicates the amount of thermal energy converted from the kinetic energy of the cluster merger (Schaal & Springel 2015; Schaal et al. 2016).

We additionally investigate the extended sample of MGS halos that have nearby MGS halos but with smaller masses. In the MGS catalog, there are 59 MGS halos with $M_{200} > 10^{14}\,M_\odot$ that have neighboring MGS halos with $M_{200} > 10^{13}\,M_\odot$ within 3 Mpc. We assume that a bow show from a high speed ($\sim$ 3000 km s$^{-1}$) merger can survive for $\sim$ 1 Gyr (e.g., the case of the Bullet cluster or El Gordo cluster; Dawson 2013; Zhang et al. 2015). During this time scale, the merging cluster can move up to $\sim$ 3 Mpc if we assume a simple linear

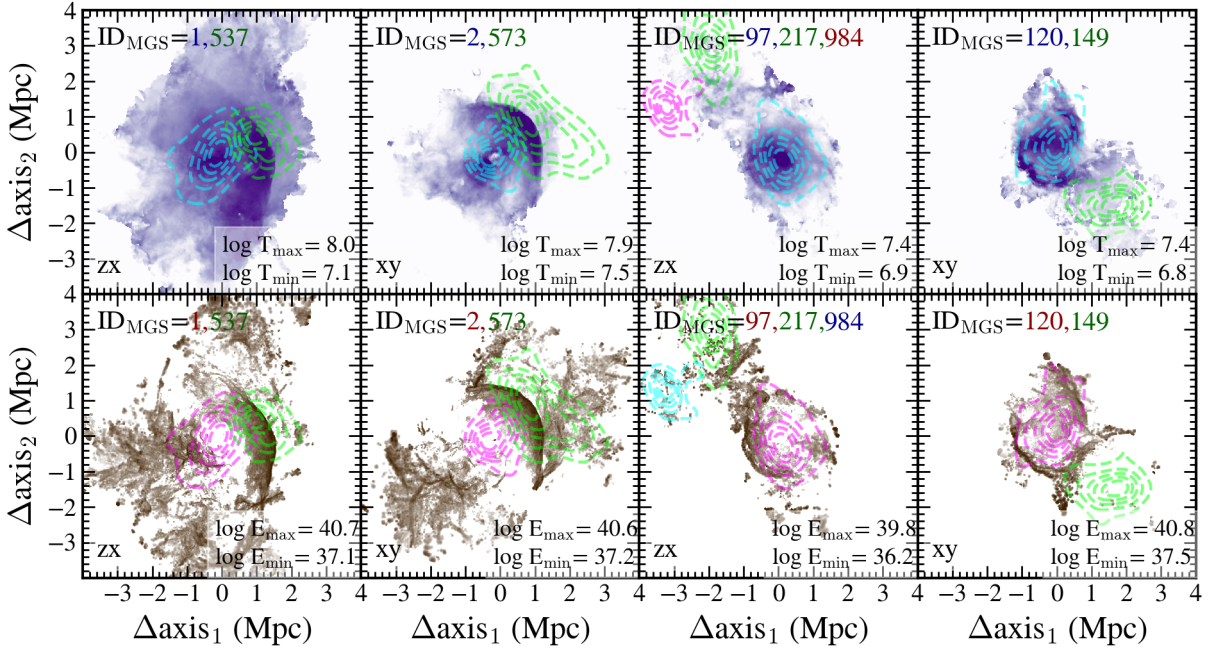

**Figure 7.** The gas temperature (upper panels) and shock-dissipated energy map (lower panels) of four FoF systems with bow shocks and multiple MGS counterparts. Dashed contours indicate the number density of member subhalos from different MGS halos. We show the distribution of subhalos on the plane (e.g., x-y, y-z, or x-z planes) where the separations of multiple structures are most distinctive. We note the minimum and maximum colormap values in units of K and erg s$^{-1}$ at the lower right of each panel. ID$_{\rm MGS}$s in each panel indicate the MGS cluster IDs. The colors of the IDs correspond to the contour colors.

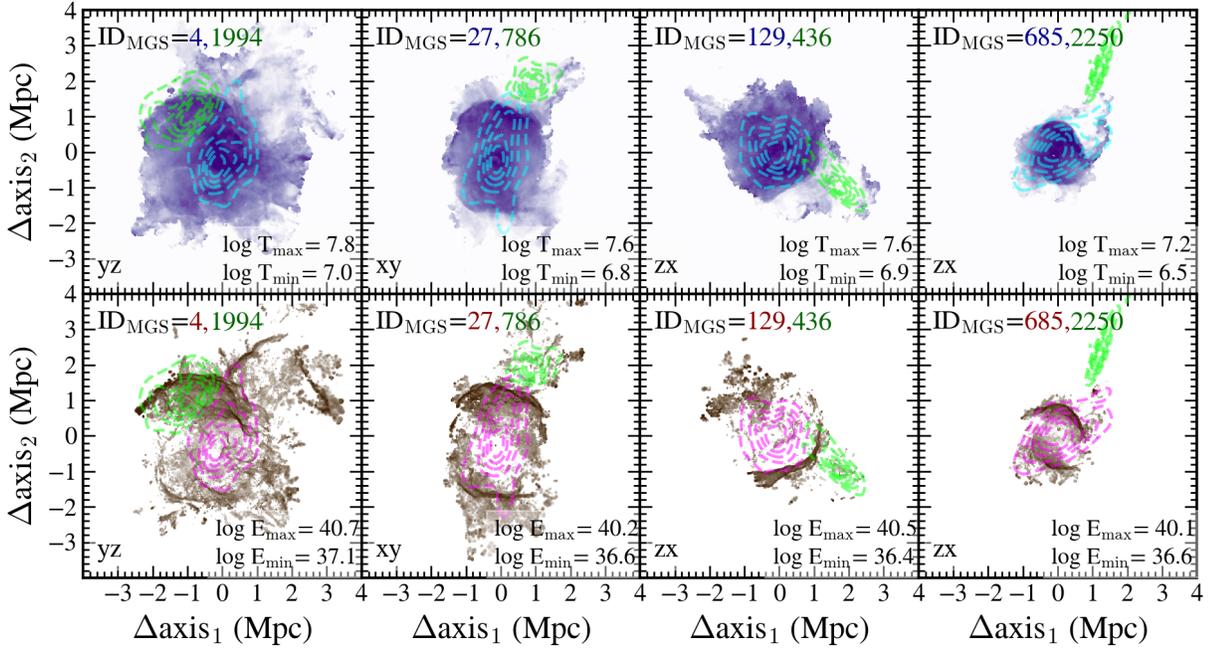

**Figure 8.** The example gas temperature (upper panels) and shock-dissipated energy map (lower panels) of MGS systems with prominent bow shocks. These MGS halos with $M_{200} > 10^{14}\,M_\odot$ have neighboring MGS halos with $M_{200} > 10^{13}\,M_\odot$ within 3 Mpc boundary. The other gas temperature map of 12 candidates with bow shock is in the Appendix.

motion. Thus, we only inspect the MGS halos that have tentative merging counterparts within 3 Mpc.

Among 59 systems, we detect 20 merging cluster candidates ($\sim 34\%$) with bow shock features in the gas temperature map. Figure 9 shows an example of four systems with prominent bow shocks in their gas temperature and shock-dissipated energy map. The gas temperature maps show the bow shock with sharp temperature gradients. The bow shock-like features appear at similar locations in the shock-dissipated energy maps. Interestingly, $\sim 85\%$ of shock fronts ($ID_{MGS}$ = 1, 2, 4, 21, 27, 31, 122, 129, 154, 233, 236, 245, 261, 372, 394, 685, and 751) are well aligned to the direction toward the neighboring MGS halo. In other words, the location of the bow shock suggests that the neighboring MGS halos probably pass through the main clusters and leave a trace.

Integrating the MGS cluster catalog with the gas temperature map analysis is a powerful tool for identifying merging clusters. To construct the merging cluster catalog efficiently, the MGS catalog can help in selecting merging cluster candidates from the large sample of clusters. Particularly, this approach will be useful for selecting merging cluster candidates with future cosmological simulations with even larger volumes (e.g., TNG-Cluster, FLAMINGO simulations; Nelson et al. 2023; Schaye et al. 2023).

## 6. CONCLUSION

We construct a new simulated cluster catalog in IllustrisTNG300 based on the Mulguisin (MGS) algorithm. Similar to the cluster identification based on spectroscopy surveys, we identify galaxy clusters by applying the MGS algorithm to the simulated subhalos. This approach is distinctive from the conventional cluster identifications in simulations that use unobservable dark matter particles. We use the linking length $b = 0.3$ (in the unit of subhalo separation), which results in the typical mass density of halos is similar to that of IllustrisTNG FoF cluster halos.

The most unique feature of the MGS cluster catalog is that the MGS algorithm separates the massive halos bundled into a single FoF halo. The multiple MGS halos classified as a single FoF halo are clearly separated in galaxy distributions and the phase-space diagram (the R-v diagram). As a result, the MGS algorithm identified 300 halos with $M_{200} > 10^{14}$ $M_{\odot}$, $\sim 10\%$ more than FoF halos in TNG300. The number difference results from the separation of multiple (potentially) independent cluster halos within $\sim 15$ FoF halos. The final MGS halo mass function is generally consistent with the FoF halo mass function.

The MGS cluster catalog has two important advantages for studying galaxy evolution and cluster formation using simulations. First, the MGS cluster catalog enables galaxy evolution studies in clusters with a well-defined cluster sample. For example, we demonstrate that the mean stellar age and the quenched fraction of MGS member subhalos clearly decreases as a function of projected distance from the MGS centers, while the similar investigation with the FoF systems is affected by the multiple dense structures within a single halo.

The MGS cluster catalog also enables an efficient search for merging cluster candidates with a small separation. Based on the visual inspection of the MGS cluster halos that have neighboring MGS halos with slightly lower mass, we identify bow-shock-like features from the gas temperature map in $\sim 34\%$ of them. In other words, the MGS cluster catalog prepares the targets for the visual inspection of the gas temperature map to identify merging clusters.

The new subhalo-based MGS cluster catalog built based on simulations is a benchmark for a direct comparison with the observed galaxy cluster catalogs. The extended cluster catalog constructed based on the future cosmological simulations with a larger volume (TNG-cluster, MilleniumTNG; Nelson et al. 2023; Hernández-Aguayo et al. 2023; Pakmor et al. 2023) will provide useful guides to understand the galaxy evolution in clusters and the evolution of clusters. A comparison of these results with future large spectroscopic surveys (e.g., DESI, 4MOST, Subaru Prime Focus Spectrograph; DESI Collaboration et al. 2016; de Jong et al. 2019; Takada et al. 2014) will enable a deeper understanding of the structure formation process.

We thank the IllustrisTNG team for providing the data. We thank Ivana Damjanov and Michele Pizzardo for their helpful discussions and comments while preparing this manuscript. L.S. and J.S. are supported by the National Research Foundation of Korea (NRF) grant funded by the Korean government (MSIT) (RS-2023-00210597). This work was supported by the New Faculty Startup Fund from Seoul National University. This work was also supported by the Global-LAMP Program of the National Research Foundation of Korea (NRF) grant funded by the Ministry of Education (No. RS-2023-00301976).

APPENDIX

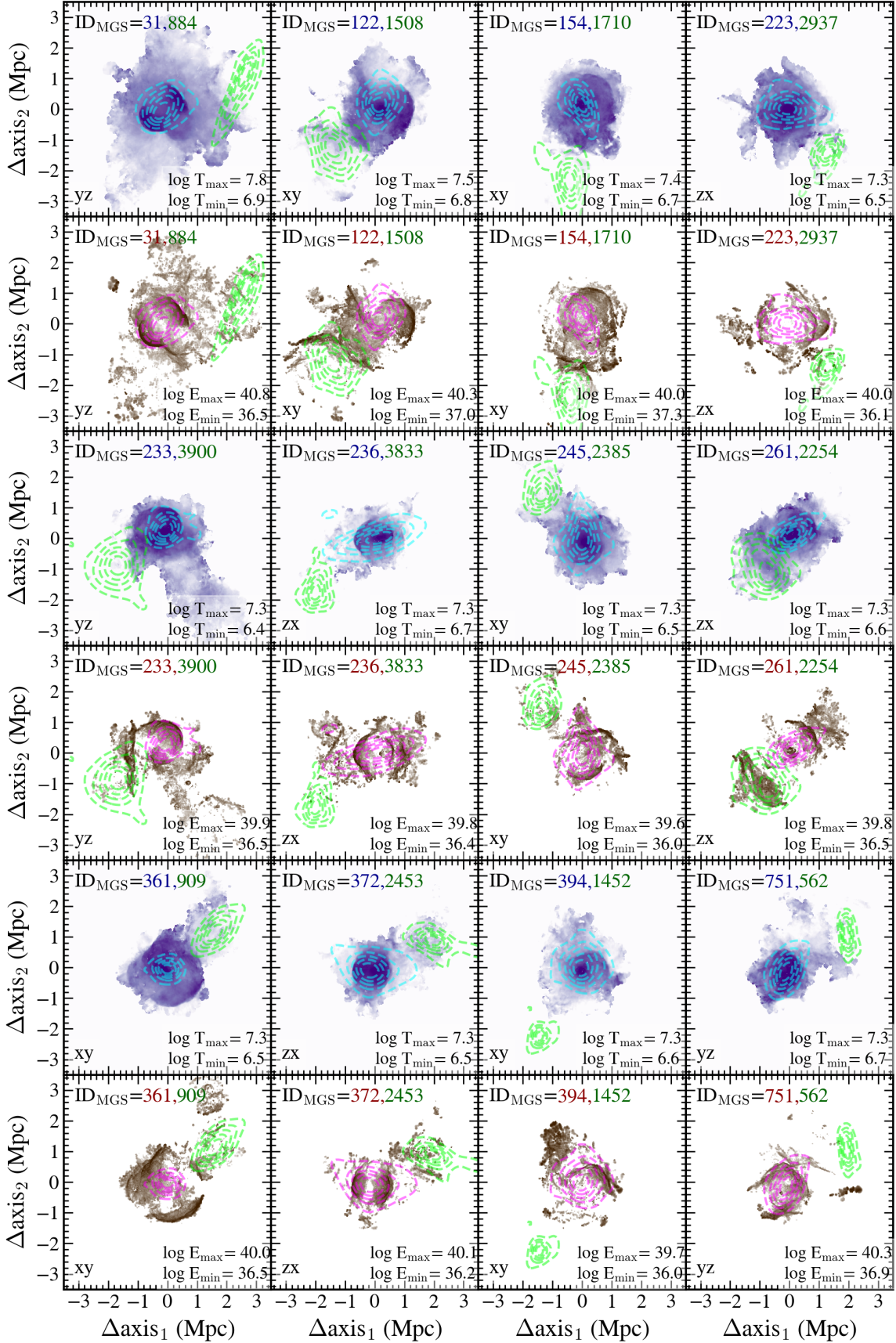

**Figure 9.** The Same as Figure 8, but for the rest merging cluster candidates with bow shocks